\documentclass{article}
\usepackage{color}
 \usepackage{amsbsy}
  \usepackage{bm}
 \usepackage{amsfonts}
\usepackage{amssymb}
\usepackage{amsmath}
\usepackage{graphics}
\usepackage{graphicx}
\begin{document}

\title{Vacuum effects on the properties of nuclear matter under an external magnetic field }
\author{R. M. Aguirre }
\date{\it{Departamento de Fisica, Facultad de Ciencias Exactas}, \\\it{Universidad Nacional de La Plata,} \\
\it{and IFLP, UNLP-CONICET, C.C. 67 (1900) La Plata, Argentina.}}


\maketitle

\begin{abstract}
The effects of the Dirac sea of the nucleons are investigated
within a covariant model of the hadronic interaction. I extend the
usual Mean Field Approximation and present a procedure to deal
with divergences which are proportional to polynomials on the
magnetic field intensity. For this purpose a nucleon propagator is
used which takes account of the full effect of the magnetic field
as well as the presence of the anomalous magnetic moments of both
protons and neutrons. I examine single-particle properties and
bulk thermodynamical quantities and conclude that within a
reasonable range of densities and magnetic intensities the effects
found are moderate.

\noindent
\\

\end{abstract}


\section{Introduction}

The interaction between matter and strong magnetic fields is a
subject of permanent research \cite{LAI,MIRANSKY}. In particular
the combination of magnetic fields and the strong interaction has
been intensively debated in the low \cite{CHAKRABARTY,
BRODERICK,DONG,RABHI,CHANDRA,BANDY,HABER, MUKHERJEE,GRASSO,SUH}
and medium energy regimes
\cite{KLEVANSKY,CHAKRABARTY2,EBERT,AVANCINI,CHAUDHURI}. In the
first case, the use of hadronic degrees of freedom is
indispensable. Among the most used models of the hadronic
interaction, the Quantum Hadro-Dynamics(QHD) has a remarkable
versatility to describe a variety of phenomena and its
results have a satisfactory accuracy when it is required.\\
QHD has been used to study the interaction of hadrons and
magnetics fields, for instance in the structure and composition of
neutron stars \cite{CHAKRABARTY, BRODERICK,DONG},the liquid-gas
phase transition \cite{RABHI}, the neutrino propagation in nuclear
matter \cite{CHANDRA}, the deconfinement phase transition
\cite{BANDY}, magnetic catalysis \cite{HABER, MUKHERJEE}, the
modification of nuclear structure \cite{GRASSO}, and the
formation of magnetic domains \cite{SUH}. \\
Within this description there is a general agreement that the
anomalous magnetic moments (AMM) of the hadrons play a
significative role when the magnetic energy approaches  the QCD
scale, i.e. $q B \approx (220 MeV)^2$
\cite{BRODERICK,DONG,CHANDRA,MUKHERJEE}.\\
One of the features of the QHD models is the simplicity of
conceptual resources and procedures. The crucial point for these
models is the Mean Field Approximation (MFA) where the meson
fields are replaced by their in medium-mean values. In addition,
the bilinear products of fermion fields are replaced by their
expectation values. In the last case the contributions coming from
the Dirac sea of fermions are usually disregarded. The procedure
is completed with the requirement of self-consistency of the
scalar meson fields, which are not directly related to conserved
charges.\\
The same procedure was adopted for a model based on the chiral
SU(3) symmetry of the strong interaction \cite{PAPAZOGLOU}, which
was used to study different aspects of hadronic matter subject
to an external magnetic field \cite{MISHRA,AGUIRR3,MISHRA1}.\\
Some attempts has been made to incorporate the vacuum contribution
within this scheme \cite{HABER, MUKHERJEE}. However, in
\cite{HABER} the AMM of the nucleons are neglected, although very
strong magnetic intensities are considered ($q\, B \approx (500
MeV)^2$). Furthermore, there is no
contribution of the neutron.\\
On the other hand, in \cite{MUKHERJEE} a low magnetic intensity
expansion is proposed for the nucleon propagator, where the
discrete energy spectrum of the protons due to the Landau
quantization is not taken into account.\\
The technical difficulties arising when the vacuum contributions
in the presence of an external magnetic field are included have
recently been considered within the Nambu and Jona-Lasinio model
of the quark interaction \cite{AVANCINI}.

An analysis of the magnitude of the vacuum effects under the
influence of strong magnetic fields, taking into account all the
physical ingredients in a coherent manner, is necessary to discuss
the validity of the usual MFA.\\
This is precisely the aim of the present work. Here a version of
the QHD model with polynomial meson interactions is used; it is
known as FSUGold \cite{TODD}. Contributions of the vacuum are
evaluated by using a nucleon propagator which includes the
anomalous magnetic moments and the full interaction with the
external magnetic field \cite{AGUIRRE1,AGUIRRE2}. This propagator
has been used to evaluate meson properties \cite{AGUIRR3,
AGUIRRE2} and the effect of the AMM within the Nambu and
Jona-Lasinio model \cite{CHAUDHURI}.\\
Within this scheme I evaluate the effective nucleon mass and
statistical properties such as the grand canonical potential and
the magnetization as functions of the baryonic density and the
magnetic intensity at zero temperature.

This work is organized as follows. In the next section the QHD
prescriptions for the MFA as well as its extension to include the
vacuum contributions are presented. Some numerical results are
discussed in Sec. III, and the last section is devoted to drawing
the conclusions of this work.

\section{Vacuum corrections to the MFA within the QHD model}

The field equations for the QHD model supplemented with the
couplings of an external magnetic field to the charge of the
proton, as well as to the anomalous magnetic moments of both
protons and neutrons, are \cite{BRODERICK}
\begin{eqnarray}
\left(i\, \not \! \partial- m + g_s \sigma + q_b \not \!\! A- g_w
\not \! \omega- g_r \bm{\tau \cdot} \not \!\!  \bm{\rho} -\kappa_b
\, \sigma_{\mu\nu} F^{\mu \nu}\right)\Psi_b=0 \label{NuclEq}
\end{eqnarray}
\begin{eqnarray}
 \left(\square + m_s^2 + g_{s2}\, \sigma + g_{s3}\,
 \sigma^2\right)\sigma=g_s \sum_b \bar{\Psi}_b \Psi_b\label{Sigma}
 \end{eqnarray}
\begin{eqnarray}
 \partial_\mu \Omega^{\mu \nu}+\left(m_w^2+ G_w \omega_\mu \omega^\mu +
 G_{r w} \bm{\rho}_\mu {\bm \cdot} \bm{\rho}^\mu\right)\omega^\nu=
 g_w \sum_b \bar{\Psi}_b \gamma^\nu\Psi_b \label{Omega}
\end{eqnarray}
\begin{eqnarray}
D_\mu R_a^{\mu \nu}+\left(m_r^2+ G_{r w} \omega_\mu
\omega^\mu\right)\rho_a^\nu=g_r \sum_{b c} \bar{\Psi}_b
\tau^{(a)}_{b c} \gamma^\nu \Psi_c \label{Rho}
\end{eqnarray}
where the index $b$ in Eq. (\ref{NuclEq}) indicates proton or
neutron, $\Omega_{\mu \nu}$ and $R_{\mu \nu}$ are the field
tensors for the $\omega$ and $\rho$ fields, and the couplings
constants are related to the notation of \cite{TODD} by
$g_s=g_{\sigma N}$, $g_{s2}=\kappa\, g_{\sigma N}^3/2$,
$g_{s3}=\lambda \,g_{\sigma N}^4/6$, $G_w=\zeta\, g_{\omega
N}^4/6$, and $G_{r w}=2\, \Lambda_w g_{r N}^2 g_{w N}^2$.

Assuming uniform matter distribution, the meson fields in these
equations are replaced by functions depending only on the bulk
properties of the system. Furthermore, the products of fermionic
fields on the right-hand side of Eqs. (\ref{Sigma}),
(\ref{Omega}), and (\ref{Rho}) are replaced by their expectation
values. Under such conditions, and adopting the reference frame of
rest matter, only the cases with $\nu=0$ gives non-zero values in
Eqs. (\ref{Omega}) and (\ref{Rho}). Finally, as the weak decay  is
not contemplated in the interaction, only the case $a=3$ in Eq.
(\ref{Rho}) gives a non-zero contribution.\\
The above mentioned expectation values can be evaluated by using
the appropriate fermion propagators
\begin{eqnarray}
{\cal N}_{s\,b}=&<\bar{\Psi}_b\,\Psi_b>&= -i\;\lim_{t'\rightarrow
t^+}\,\text{Tr}\{G_b(t,\vec{r},t',\vec{r})\}
 \\
{\cal N}^\nu_b=&<\bar{\Psi}_b\,
\gamma^\nu\,\Psi_b>&=-i\;\lim_{t'\rightarrow
t^+}\,\text{Tr}\{\gamma^\nu\,G_b(t,\vec{r},t',\vec{r})\}
\end{eqnarray}
In the momentum representation they can be rewritten as
\begin{eqnarray}
{\cal N}_{s\,b}&=& -i\;\lim_{\epsilon \rightarrow 0^+}\,\int
\frac{d^4p}{(2\pi)^4} e^{-i p_0 \epsilon}\;\text{Tr}\{G_b(p)\}
\label{SDen} \\
{\cal N}^\nu_b&=&-i\;\lim_{\epsilon \rightarrow 0^+}\,\int
\frac{d^4p}{(2\pi)^4} e^{-i p_0 \epsilon}\;\text{Tr}\{\gamma^\nu
G_b(p)\} \label{0Den}
\end{eqnarray}

 In \cite{AGUIRRE1} a fermion propagator which includes the
full interaction with the external magnetic field, through its
coupling to the proton charge and to the AMM also, was used for
this purpose. For the sake of completeness the explicit form of
the neutron propagator is
\begin{eqnarray}
G_n(x',x)= \sum_s \int \frac{d^4p}{(2 \pi)^4} e^{-i p^\mu\,(x_\mu
'-x_\mu)} \Lambda_s \; \Xi_p   \label{PropN}\end{eqnarray}
 where
\begin{eqnarray}
\Lambda_s&=&\frac{ s}{2 \Delta}i\; \gamma^1 \gamma^2\left[ \not \!
u+ i \gamma^1 \gamma^2 (s \Delta-\kappa B)\right] \left( \not \!
v+m+ i s \Delta \gamma^1 \gamma^2\right)  \\
\Xi_p&=&\frac{1}{p_0^2-E_s^2+i\epsilon}+ 2
\pi\,i\,n_F(p_0)\,\delta(p_0^2-E_s^2) \label{DecomN}
\end{eqnarray}
whereas for the proton one has
\begin{equation}
G_p(x',x)=e^{i \Phi } \int \frac{d^4 p}{(2 \pi)^4} e^{-i
p^\mu\,(x'_\mu-x_\mu)} \left[ G_0(p)+e^{-p_\bot^2/\beta}
\sum_{n,s}(-1)^n  G_{n,s} (p)\right] \label{PropP}
\end{equation}
where
\begin{eqnarray}
G_0(p)&=&2 e^{-p_\bot^2/\beta}\Lambda^0 \; \Xi_{0\, 1}
\end{eqnarray}
\begin{eqnarray}
G_{n s}(p)&=&\frac{\Delta_n+s m}{2 \Delta_n}\Big\{( \not \!
u-\kappa_p B+s \Delta_n) \left(1+i \gamma^1 \gamma^2\right) L_n(2
p_\bot^2/\beta)-( \not \! u+\kappa_p B-s \Delta_n)
\nonumber\\
&&\times \left(1-i \gamma^1 \gamma^2\right) \frac{s \Delta_n-m}{s
\Delta_n+m} L_{n-1}(2 p_\bot^2/\beta)+ \left( \not \! u\, i
\gamma^1 \gamma^2+ s \Delta_n-\kappa_p B\right)  \not \! v \frac{s
\Delta_n- m}{p_\bot^2}
\nonumber \\
&&\times \left[ L_n(2 p_\bot^2/\beta)-L_{n-1}(2
p_\bot^2/\beta)\right]\Big\}
\; \Xi_{n s}  \\
\Xi_{n s}&=&\frac{1}{p_0^2-E_{n s}^2+i\epsilon}+2
\pi\,i\,n_F(p_0)\,\delta(p_0^2-E_{n s}^2) \label{DecomP}
\end{eqnarray}

In these expressions the index $s=\pm 1$ corresponds to the
projection of the spin in the direction of the uniform magnetic
field, the index $n\geq 1$ takes account of the discrete Landau
levels, and the following notation is used: $\beta=q B$, $\not \!
u=p_0 \gamma^0-p_z \gamma^3$, $\not \!   v=-p_x \, \gamma^1-p_y\,
\gamma^2$, $p_\bot^2=p_x^2+p_y^2$, $L_m$ stands for the Laguerre
polynomial of order $m$, and
\begin{eqnarray}
E_s&=&\sqrt{p_z^2+(\Delta-s\,\kappa_n \nonumber
B)^2}\nonumber\\
 \Delta&=&\sqrt{m^2+p^2_x+p^2_y} \nonumber \\
E_{n s}&=&\sqrt{p_z^2+(\Delta_n-s\,\kappa_p B)^2}\nonumber \\
\Delta_n&=&\sqrt{m^2+2 n q B} \nonumber
\end{eqnarray}
Finally, the phase factor $\Phi=q B(x+x')(y'-y)/2$ embodies the
gauge fixing.

If these propagators are used in Eqs. (\ref{SDen}) and
(\ref{0Den}), but keeping only the second terms of Eqs.
(\ref{DecomN}) and (\ref{DecomP}), then the MFA is obtained
\cite{AGUIRRE1}. The correction to the densities coming from the
Dirac sea of nucleons can be evaluated by using Eqs. (\ref{SDen})
and (\ref{0Den}) but retaining only the first terms of Eqs.
(\ref{DecomN}) and (\ref{DecomP}). The expressions thus obtained
are divergent and must be renormalized. Since the main residue in
a Lorenz expansion depends on the magnetic intensity, a
regularization procedure must be defined to extract relevant
contributions. The details of this calculations are left for the
Appendix, and here the final results are shown. ${\cal
N}_\nu^{\text{vac}}=0$ for protons and neutrons,
\begin{eqnarray}
{\cal N}_{s\,p}^{\text{reg}}&=&\frac{1}{4 \pi^2}\Big[ 2 \beta
\kappa_p B+ \frac{\beta^2}{3 m}-m^3 + 2 \beta (m+\kappa_p B)
\ln\left(\frac{m}{m+\kappa_p B}\right)\nonumber \\
&&+m\left(\beta-m^2\right) \ln\left(\frac{2 \beta}{m^2}\right)-2 m
\beta \ln\left(\frac{\Gamma(m^2/2\beta)}{\sqrt{2 \pi}}\right)\Big]
\label{VacP}
\end{eqnarray}
for protons, and
\begin{eqnarray}
{\cal N}_{s\,n}^{\text{reg}}=\frac{m}{4 \pi^2}\left[6 (\kappa_n
B)^2+(m-\kappa_n B)^2 \ln\left(\frac{m}{m-\kappa_n
B}\right)+(m+\kappa_n B)^2 \ln\left(\frac{m}{m+\kappa_n B}\right)
\right] \label{VacN}
\end{eqnarray}
for neutrons.\\
It must be pointed out that the first three terms on the
right-hand side of Eq. (\ref{VacP}) come from the subtraction
proposed in the regularization procedure. It is interesting to
note that Eq.(\ref{VacN}) becomes zero if $\kappa_n=0$ is taken,
while, taking $\kappa_p=0$ in Eq. (\ref{VacP}) and writing
$x=m^2/2\beta$, this equation reduces to
\begin{eqnarray}
-\frac{m \beta}{2 \pi^2}\Big[ -\frac{1}{12 x}+x -x \ln(x)
+\frac{1}{2} \ln\left(\frac{x}{2 \pi}\right)+
\ln\left(\Gamma(x)\right)\Big]. \label{Haber}
\end{eqnarray}
Here the first two terms correspond to the second and third terms
of (\ref{VacP}). With exception of the first term between square
brackets in Eq.(\ref{Haber}), this expression can be recognized as
the vacuum correction term on the right-hand side of Eq. (28a) of
Ref. \cite{HABER}.  As explained before, the discrepant term is
justified by the subtraction prescription used here.

With these results, one can extend the standard definition of the
effective nucleon mass in QHD models $m=m_0-g_s \, \bar{\sigma}$,
where $m_0=939$ MeV and the uniform mean value $\bar{\sigma}$ is
obtained from Eq.(\ref{Sigma}) by neglecting the coordinate
dependence and replacing $\bar{\Psi}_b \Psi_b \rightarrow {\cal
N}_{s \, b}$. And the scalar baryonic density is the sum of the
MFA result and the vacuum correction given by Eqs. (\ref{VacP})
and (\ref{VacN}). The self-consistency is imposed by evaluating
${\cal N}_{s \, b}$ in the unknown mass $m$.

For further applications it is useful to obtain the vacuum
correction to the energy density. The baryonic contribution to the
energy density arises from the mean field value of the Hamiltonian
density operator \cite{AGUIRRE1}
\begin{eqnarray}
{\cal E}_b=<{\cal H}_b>=-i\;\lim_{t'\rightarrow
t^+}\,\text{Tr}\left\{i \gamma^0 \frac{\partial}{\partial
t}\,G_b(t,\vec{r},t',\vec{r})\right\}. \label{EDen}
\end{eqnarray}
By using the method described in the Appendix, the following
results are obtained:
\begin{eqnarray}
{\cal E}_n^{\text{reg}}&=&\frac{1}{48 \pi^2}\Big\{\left[(\kappa_n
B)^4-6 m^2(\kappa_n B)^2-3 m^4 \right]\ln\left[\frac{m^2-(\kappa_n
B)^2}{m^2}\right]-4 \kappa_n B m^3 \ln\left(\frac{m+\kappa_n
B}{m-\kappa_n B}\right)\nonumber \\
&&+\frac{13}{6}(\kappa_n B)^2\left[6 m^2-(\kappa_n
B)^2\right]\Big\} \label{Evacn}\\
{\cal E}_p^{\text{reg}}&=&\frac{1}{8 \pi^2}\Big\{-4 \beta^2
\zeta'(-1,\lambda)-\frac{1}{2}\ln\left(\frac{m^2}{2\beta}\right)\left(\mu^2-2
\beta
\mu+\frac{2}{3}\beta^2\right)-\frac{1}{4}m^4+\left[\beta+(\kappa_p
B)^2\right]^2\nonumber \\
&&-2 \beta (m+\kappa_p B)^2 \ln\left(\frac{m+\kappa_p
B}{m}\right)+\frac{1}{3}\beta(\beta+2 m \kappa_p
B)\left(\frac{\kappa_p B}{m}\right)^2 -\frac{2}{3}\beta^2+2 m
\beta \kappa_p B \nonumber \\
&&+\frac{1}{45}\left(\frac{\beta}{m}\right)^4\Big\} \label{Evacp}
\end{eqnarray}
where $m$ stands for the vacuum value, i.e., $m=m_0$,
$\mu=m_0^2+(\kappa_p B)^2$, $\lambda=\mu/2\beta$, and $\zeta'$
indicates the derivative of the Hurwitz zeta function
respect to its first argument.\\
The magnetization of the system can be evaluated as ${\cal
M}=\left(\partial {\cal E}/\partial B\right)_{N_b}$, where ${\cal
E}$ is the hadronic contribution to the total energy
\cite{BRODERICK}. Finally, using the chemical potentials
associated with the conservation of the baryonic number of protons
and neutrons, the pressure at zero temperature can be evaluated as
$P=\sum_b \mu_b n_b-{\cal E}$.

\section{Results and discussion}

In this section several properties of dense nuclear matter are
analyzed, considering baryonic densities lesser than three times
the normal nuclear density and magnetic intensities between
$10^{14}$  and $10^{19}$ G. The isospin composition of matter has
also been taken as a relevant  variable to be examined. However,
the main conclusions of this work are basically independent of the
isospin asymmetry, so I examine in what follows the
symmetric nuclear matter case. \\
The parameters are taken from the FSU model \cite{TODD}.

First the effective nucleon mass is analyzed. It is directly
affected by the vacuum corrections to the scalar densities given
by Eq. (\ref{SDen}). In Fig. 1 the dependence of $m$ on the
magnetic intensity is  displayed at constant baryonic density. For
this purpose I take $n/n_0=0, \, 1, \, 2$, where $n_0$ stands for
the normal nuclear density. In each case the results including
vacuum correction (CC)  and without it (NC) are
 compared. For intensities below $5 \times 10^{18}$ G, both cases yields
very similar results. Up to this point, the inclusion of
corrections produces higher values of the effective mass. This
effect is more pronounced at lower densities, for instance at
$B=10^{19}$ G the differences between the two cases are 2.7, 2.2,
and 1.9 MeV for the densities $n/n_0=0, \,1, \, 2$ respectively.
This can be understood because the vacuum effect at fixed magnetic
intensity reduces to a constant which dominates at very low
densities. But, as the density increases, the MFA provides
a growing contribution.\\
In Fig. 1 a wider range of magnetic intensities is considered in
order to compare with previous results. For instance in
\cite{HABER} the difference for the nucleon mass between the CC
and NC cases at zero density is approximately $10$ and $20$ MeV
for $B=10^{19}$  and $1.6 \times 10^{19}$ G, respectively, whereas
in these calculations I have obtained $3$ MeV and $9$ MeV for the
same values of the magnetic intensity. This fact is illustrated in
Fig. 2a, where the results of the present work (solid line) are
contrasted with those obtained by following the procedure and
model parameters used in \cite{HABER} (dotted line). In order to
expand the analysis, I also show the results obtained with the FSU
model including the AMM  but adopting the regularization procedure
of Ref.\cite{HABER} (dash-dotted line). I conclude that the
numerical discrepancy comes mainly from the different
regularization prescriptions and in a minor degree can be ascribed
to the model parameters and to the presence of the AMM.
Notwithstanding, even for the extreme intensity $B=10^{19}$ G, all
the approaches predict an increment of the effective mass not
greater than $2 \%$ of the experimental value $m_0$.\\
The role of the anomalous magnetic moments is analyzed throughout
the three panels of Fig. 2. The outcome  for the present
calculations with $\kappa_p=\kappa_n=0$ at fixed density is
represented by the curves with dashed lines. Neglecting the AMM
yields a decreasing effective mass, that increasingly differs from
the full calculations as the magnetic intensity and the baryonic
density are increased. For a given density and low intensities ($B
< 10^{18}$ G) the results with or without AMM are almost
identical, whereas for the greatest intensity examined here
($B=1,6\times 10^{19}$ G)
 the difference grows from $10$ to $25$ MeV.

As the next step, the pressure at zero temperature is studied. In
Fig.3 the pressure as a function of the particle density is shown
for a constant magnetic intensity, for the specific values
$B=10^{18}, \, 5\times 10^{18}$ and $10^{19}$ G. Again a
comparison between the CC and NC cases is made. For each pair of
curves the CC case presents higher values for the whole range of
densities. Furthermore the separation between each pair does not
vary significatively with the density. This can be explained
because the vacuum correction does not depend on the density, and
the changes induced in the meson mean values are so weak that the
CC curve practically copies the same features of the NC one.
However the shift between the twin curves increases appreciably
with the magnitude of $B$.  This behavior remains when the
isospin composition is varied.\\
An interesting consequence which can be appreciated in Fig. 3, is
that the vacuum correction preserves the thermodynamical
instabilities of the MFA. Therefore a spinodal decomposition
similar to that shown in \cite{RABHI} must be expected, with the
same range of densities but extending to higher pressures.\\
It must be said that in order to allow an easier comparison within
the same figure, the constant contribution of the magnetic field
to the total energy density is not shown in Fig. 3.

As the last statistical subject to be analyzed the magnetization
${\cal M}$ induced by the external magnetic field is  considered.
It must be mentioned that within the scheme of regularization
presented here, one can obtain finite vacuum contributions to
${\cal M}$ as also would be
 the case for the magnetic susceptibility $\chi=\partial{\cal M}/\partial
B$.\\
 It is known
that ${\cal M}$ is a very weak quantity, since it is proportional
to the electric charge. The correction to the energy density,
given by Eqs. (\ref{Evacn}) and (\ref{Evacp}), does not depend on
the matter density, nor does its contribution to the
magnetization. Hence I analyze the dependence on $B$ of the
difference $\Delta {\cal M}$ between the full magnetization and
the corresponding result without vacuum effect. To compare with
previous calculations \cite{DONG, RABHI2} which used the same
hadronic interaction, the adimensional ratio $\Delta{\cal M}/q^2B$
as a function of the magnetic intensity is  shown in Fig. 4,
separately for the proton and neutron contributions within the
range $10^{17}$  $< B < 10^{19}$ G. By considering the
 results of \cite{RABHI2} for $B>10^{18}$ G, an almost general
 trend is that this quantity decreases by increasing the magnetic
 intensity and decreasing the matter density. Due to the tiny
 results I obtained for the neutron component, one can expect
that the vacuum corrections to this component could have
significative effects only in the very low density regime.
However, in this regime the assumption of homogeneous matter is
not valid. So, one can conclude that the vacuum correction to the
neutron component is negligible, with the possible exception of
certain special configurations, for instance, in the case of the
neutron gas surrounding the nuclear clusters in the inner crust of
a neutron star.\\
In regard to the proton component, a growing magnitude is obtained
as $B$ is increased, reaching at $B=10^{19}$ G the value
$10^{-5}$. This represents approximately  $10 \%$ of the result at
a density $n/n_0 = 0.2$ (see Fig. 8 of Ref. \cite{RABHI2}) and $1
\%$ at $n/n_0 = 1.2$. In conclusion, the vacuum correction for the
proton component starts to be significant for intensities $B
\approx 10^{19}$ G, modifying the MFA result only by a few percent
 for densities below the normal nuclear density.

\section{Conclusions}

In this work I have proposed an extension of the MFA for nuclear
matter under the effect of a uniform magnetic field, by including
contributions from the Dirac sea of hadrons. I have used the
covariant FSU model of the nuclear interaction \cite{TODD} and the
calculations have been made by using a covariant propagator which
takes account of the full effect of the magnetic field as well as
the effect of the anomalous magnetic moments. Hence, several
issues left open in previous investigations \cite{HABER,
MUKHERJEE} are considered.

Since the interaction used is just an effective model of the
strong interaction, I have not considered a renormalization
scheme. In particular I did not try to renormalize the external
magnetic field since, within the model used, it is not a dynamical
variable. I have proposed instead a regularization procedure to
obtain physically
meaningful results from the divergent contributions.\\
The procedure has the advantage of yielding finite results for the
vacuum correction to the magnetization ${\cal M}=\partial {\cal
E}/\partial B$ as well as to the higher order derivatives, for
instance the magnetic susceptibility $\chi=\partial {\cal
M}/\partial B$.

Within the scheme proposed I have evaluated different nuclear
properties at zero temperature. The effective nucleon mass is
representative of the single-particle properties,  and the
pressure and the magnetization correspond to bulk properties in
thermodynamical equilibrium. They have been analyzed for a range
of matter densities and magnetic intensities that ensures
the confidence in the model.\\
For all the cases I have obtained moderate corrections, which
becomes significant for densities below the normal nuclear
density, and magnetic intensities above $10^{18}$ G. \\
Taking into account that QHD models use hadronic degrees of
freedom exclusively and that their parameters are adjusted to the
low energy phenomenology, it is consistent that vacuum corrections
do not reveal high energy manifestations. These conclusions
support the validity of the MFA for the regime of parameters
studied.

\section{Acknowledgements} This work has been partially supported
by a grant from the Consejo Nacional de Investigaciones
Cientificas y Tecnicas,  Argentina.

\appendix
\section{Regularization of the vacuum contribution to the nuclear densities}

I start with the nuclear current (\ref{0Den}). By using the vacuum
component of either Eq.(\ref{PropN}) or Eq. (\ref{PropP}), it is
found that the integrand of Eq. (\ref{0Den}) is odd in the
integration variables, hence, by symmetric integration it is zero
for all $\nu$.

 Next, the neutron scalar density Eq.(\ref{SDen}) is considered.
 Using the neutron propagator, one finds for the vacuum term
\begin{eqnarray}
{\cal N}_{s\,n}^{\text{vac}}&=&-\frac{2 \,i }{(2\,
\pi)^4}\frac{m}{\nu}\, \sum_s \int\; d^4p\; \frac{\Delta-s\kappa_n
B}{\Delta}\, \frac{\nu}{u_p^2-(\Delta-s\kappa_n B)^2+i
\varepsilon}
\end{eqnarray}
where a regularization parameter $\nu$ with dimension of squared
mass has been introduced. After passing to a bidimensional
Euclidean space in the variables $p_0$ and $p_z$, this can be
written as
\begin{eqnarray}
\frac{-2}{(2\, \pi)^4}\frac{m}{\nu}\, \sum_s \int_0^\infty d\tau
\int\, d^2p_\bot \frac{\Delta-s\kappa_n B}{\Delta}\,
\exp\left[-\tau(\Delta-s \kappa_n B)^2/\nu\right] \int d^2p_E
\;e^{- \tau p_E^2/\nu}
\end{eqnarray}
After performing the momentum integrals, one obtains
\begin{eqnarray}
-\frac{m\, \nu}{8 \pi^2}\lim_{\epsilon \rightarrow 0}\sum_s
\int_0^\infty d\tau\; \tau^{\epsilon-2} e^{-\tau (m-s\kappa_n
B)^2/\nu}
\end{eqnarray}
which by the change of variable $t=\tau (m-s \kappa_n B)^2/\nu$,
takes the form
\begin{eqnarray}
-\frac{m}{8 \pi^2}\lim_{\epsilon \rightarrow
0}\Gamma(\epsilon-1)\;\sum_s (m-s\kappa_n B)^2 \left[
\frac{(m-s\kappa_n B)^2}{\nu}\right]^{-\epsilon}
\end{eqnarray}
Here a single pole can be distinguished from the finite
contributions
\begin{eqnarray}
\frac{m}{8 \pi^2}\lim_{\epsilon \rightarrow 0}\left\{
\left(\frac{1}{\epsilon}+1-\gamma\right)2\left(m^2+\kappa_n^2B^2\right)-\sum_s\left(m-s\kappa_n
B\right)\log\left[\frac{(m-s\kappa_n B)^2}{\nu}\right]+{\cal
O}(\epsilon)\right\} \label{RegSdenN}
\end{eqnarray}
when $B \rightarrow 0$ this expression reduces to the standard
result, obtained for instance by dimensional regularization from
the Feynman propagator.\\
The residue of this pole depends quadratically on $B$, hence I
propose a regularization procedure to extract finite
contributions. Before taking the limit, I subtract from
Eq.(\ref{RegSdenN}) its Taylor expansion in $B$ of order 2,
evaluated at zero baryonic density:
\begin{eqnarray}
{\cal N}_{s\,n}^{\text{reg}}&=&{\cal
N}_{s\,n}^{\text{vac}}-\sum_{j=0}^2 \Big[\frac{\partial^j {\cal
N}_{s\,n}^{\text{vac}}}{\partial B^j}\Big]_0 \frac{B^j}{j!}
\label{RegDef}
\end{eqnarray}
Choosing $\nu=m^2$ the results shown in Eq.(\ref{VacN}) is
obtained.

In the next step I consider the proton scalar density of
Eq.(\ref{SDen}) using Eq.(\ref{PropP}):
\begin{eqnarray}
{\cal N}_{s\,p}^{\text{vac}}&=&-\frac{2 \,i }{(2\, \pi)^4}
\sum_{n\, s}  \frac{(-1)^n}{ \Delta_n}\left(s\,\Delta_n-\kappa_p
B\right)\int\; d^4p\;e^{-p_\bot^2/\beta} \Big[ \left(\Delta_n+s\,
m\right)L_n\left(2
p_\bot^2/\beta\right)\nonumber \\
&&+\left(\Delta_n-s\, m\right)L_{n-1}\left(2
p_\bot^2/\beta\right)\Big] \frac{\nu}{u_p^2-(\Delta_n-s\kappa_p
B)^2+i \varepsilon}
\end{eqnarray}
where terms that are null by symmetric integration have been
disregarded.\\
By following the same first steps described previously, one
arrives at
\begin{eqnarray}
-\frac{1}{(2\, \pi)^3} \sum_{n\, s}  \frac{(-1)^n}{
\Delta_n}\left(s\,\Delta_n-\kappa_p B\right)\int_0^\infty
\frac{d\tau}{\tau} \exp\left[-\tau\left(\Delta-s \kappa_p
B\right)^2/\nu\right] \int\; d^2p_\bot \;e^{-p_\bot^2/\beta}
\nonumber \\ \times \Big[ \left(\Delta_n+s\, m\right)L_n\left(2
p_\bot^2/\beta\right) +\left(\Delta_n-s\, m\right)L_{n-1}\left(2
p_\bot^2/\beta\right)\Big]
\end{eqnarray}
The remaining momentum integration can be performed with the help
of formula 7.414 6 of \cite{G&R}, giving
\begin{eqnarray}
\frac{\beta}{(2\pi)^2}\int_0^\infty
\frac{d\tau}{\tau}\left[\left(m+\kappa_p
B\right)e^{-\tau(m+\kappa_p B)^2/\nu}- m \sum_{s, \,n=0}
\frac{\Delta_n-s \kappa_p B}{\Delta_n}
\;e^{-\tau(\Delta_n-s\kappa_p B)^2 /\nu}\right]
\end{eqnarray}
With the aim of isolating the divergent term, this equation can be
rewritten as
\begin{eqnarray}
\frac{\beta}{(2\pi)^2} \lim_{\epsilon \rightarrow 0}\nu^{\epsilon}
\int_0^\infty dt\; t^{\epsilon-1} e^{-t}\left[\left(m+\kappa_p
B\right)\left(m+\kappa_p B\right)^{-2\epsilon}- m \sum_{s, \,n=0}
\frac{\Delta_n-s \kappa_p B}{\Delta_n} \;\left(\Delta_n-s\kappa_p
B\right)^{-2\epsilon} \right]\nonumber \\
=\frac{\beta}{(2\pi)^2} \lim_{\epsilon \rightarrow
0}\nu^{\epsilon} \Gamma(\epsilon)\left\{\left(m+\kappa_p
B\right)\left(m+\kappa_p B\right)^{-2\epsilon}-m
\sum_{n=0}\Delta_n^{-2\epsilon}\left[\left(1-\frac{\kappa_p
B}{\Delta_n}\right)^{1-2\epsilon}+\left(1+\frac{\kappa_p
B}{\Delta_n}\right)^{1-2\epsilon}\right]\right\}\nonumber
\end{eqnarray}
In order to write this in a compact form it must be noted that in
the last line of this equation, the term between square brackets
reduces to 2 for $\epsilon=0$. So I approximate it by the
following expression:
\begin{eqnarray}
\frac{\beta}{(2\pi)^2} \lim_{\epsilon \rightarrow 0}\nu^{\epsilon}
\Gamma(\epsilon)\left[\left(m+\kappa_p B\right)\left(m+\kappa_p
B\right)^{-2\epsilon}-\frac{2 m}{(2\beta)^\epsilon}
\sum_{n=0}\frac{1}{(n+m^2/2\beta)^\epsilon}\right].
\end{eqnarray}
The sum over the Landau levels is the definition of the Hurwitz
zeta function $\zeta(\epsilon,m^2/2\beta)$ which can be extended
analytically to
the whole complex plane.\\
The formula 9.533 3 of \cite{G&R} is used to isolate a single pole
in the last equation:
\begin{eqnarray}
\frac{\beta}{(2\pi)^2} \lim_{\epsilon \rightarrow 0}\Big\{
\left(\frac{1}{\epsilon}-\gamma\right)\left[m+\kappa_p B-2 m
\zeta(0,m^2/2\beta)\right]-(m+\kappa_p
B)\log\left[\frac{(m+\kappa_p B)^2}{\nu}\right]\nonumber \\
-2m\left[\log\left(\Gamma(m^2/2\beta)\right)-\frac{1}{2}\log(2\pi)+\frac{1}{2}\left(1-\frac{m^2}{\beta}\right)\log\left(\frac{\nu}{2
\beta}\right)\right]+{\cal O}(\epsilon)\Big\}. \label{what1}
\end{eqnarray}
Using the fact that $\zeta(0,x)=-x+1/2$, it can be seen that the
residue of the pole depends on $B$ through a polynomial of order
2. Therefore the regularized proton scalar density can be defined
in a similar way as in Eq.(\ref{RegDef}). Thus the results shown
in
Eq. (\ref{VacP}) are obtained by adopting $\nu=m^2$.\\
As in the neutron case, when $B\rightarrow 0$ in Eq. (\ref{what1})
it yields the same result as obtained by using the normal Feynman
propagator in Eq.(\ref{SDen}).

Finally, Eq.(\ref{EDen}) is treated by passing to the momentum
representation
\begin{eqnarray}
{\cal E}_b=-i\;\lim_{\epsilon\rightarrow 0^+}\,\int\,
\frac{d^4p}{(2\pi)^4} e^{-i \epsilon p_0} p_0 \text{Tr}\left\{
\gamma^0 \,G_b(p)\right\}. \label{EDenP}
\end{eqnarray}
Inserting the Feynman term of the neutron propagator, it yields
\begin{eqnarray}
&&{\cal E}_n^{\text{vac}}=-  \frac{2 i}{(2 \pi)^4}\sum_s \int \,
d^4p \frac{p_0}{u_p^2-(\Delta-s\kappa_n B)^2+i \varepsilon}
\nonumber
\\
&=&\frac{\nu^2}{(4 \pi)^2}\Big\{\sum_s\int_0^\infty
\frac{d\tau}{\tau^3}\, e^{-\tau(m-s\kappa_n
B)^2/\nu}+2\,\frac{\kappa_n B}{\nu} \int_{m-\kappa_n
B}^{m+\kappa_n B} dz\,\int_0^\infty \frac{d\tau}{\tau^2}\;e^{-\tau
z^2/\nu}\Big\}\nonumber
\end{eqnarray}
where the second line was obtained after introducing an
exponential form for the denominator, passing to the Euclidean
space in $p_0$ and $p_z$ coordinates, and performing the momentum
integrals. In the second term between curly brackets there is a
remnant of the $p_\bot$ integration that can be solved in terms of
the incomplete error function. However I prefer to keep this form
to simplify the following steps.\\
To isolate divergent terms, it can be written in the form
\begin{eqnarray}
\frac{\nu^2}{(4 \pi)^2}\lim_{\epsilon \rightarrow
0}\Big\{\sum_s\left[\frac{(m-s\kappa_n
B)^2}{\nu}\right]^{2-\epsilon}\int_0^\infty dt\,
t^{\epsilon-3}\,e^{-t}+2\,\frac{\kappa_n B}{\nu} \int_{m-\kappa_n
B}^{m+\kappa_n B}
dz\,\left(\frac{z^2}{\nu}\right)^{1-\epsilon}\int_0^\infty
dt\,t^{\epsilon-2}\;e^{-t}\Big\}\nonumber
\end{eqnarray}
Integrations over $t$ can be identified with the gamma functions
$\Gamma(\epsilon-2)$ and $\Gamma(\epsilon-1)$.  After a Lorenz
expansion, this can be rewritten as
\begin{eqnarray}
\frac{\nu^2}{(4 \pi)^2}\lim_{\epsilon \rightarrow 0}\Big\{
\left(\frac{1}{\epsilon}+\frac{3}{2}-\gamma\right)\left(m^4-\frac{1}{3}\kappa_n^4B^4+2
m^2 \kappa_n^2B^2\right)+\frac{10}{9}\kappa_n^2B^2\left(3
m^2+\kappa_n^2B^2\right) \nonumber \\
-\left(m^4-\frac{1}{3}\kappa_n^4B^4+2 m^2
\kappa_n^2B^2\right)\ln\left(\frac{m^2-\kappa_n^2B^2}{\nu}\right)-\frac{8}{3}\kappa_n
B m^3 \ln\left(\frac{m+\kappa_n B}{m-\kappa_n B}\right)\Big\}.
\label{what2}
\end{eqnarray}
For $B \rightarrow 0$ and $\nu=m^2$ the right behavior
\begin{eqnarray}
\frac{m^4}{8\pi^2}\left(\frac{1}{\epsilon}+\frac{3}{2}-\gamma\right)
\label{EB0Lim}
\end{eqnarray}
is obtained.\\
The residue is a polynomial of order 4 in B, so I propose
\begin{eqnarray}
{\cal E}_n^{\text{reg}}={\cal E}_n^{\text{vac}}-\sum_{j=0}^4
\left[\frac{\partial^j{\cal E}_n^{\text{vac}}}{\partial
B^j}\right]_0\frac{B^j}{j!}. \label{ERegProc}
\end{eqnarray}
Applying this prescription to Eq.(\ref{what2}) yields the results
shown in Eq.(\ref{Evacn}).

For the proton case, Eq.(\ref{EDenP}) yields
\begin{eqnarray}
{\cal E}_p^{\text{vac}}=-\frac{2\,
i}{(2\pi)^4}\sum_{s,n}\frac{(-1)^n}{\Delta_n} \int
d^4p\,\frac{p_0^2\; e^{-p_\bot^2/\beta}}{u_p^2-(\Delta_n-s\kappa_p
B)^2+i \varepsilon} \left[\left(\Delta_n+s\,
m\right)L_n-\left(\Delta_n-s\, m\right)L_{n-1}\right] \nonumber
\end{eqnarray}
where the argument of the Laguerre functions $L_k$ is
$2\,p_\bot^2/\beta$. By the same procedure described previously,
the following expression is obtained:
\begin{eqnarray}
\frac{\nu \beta}{8\pi^2}\int_0^\infty
\frac{d\tau}{\tau^2}\left[\sum_{s,n=0}e^{-\tau (\Delta_n-s\kappa_p
B)^2/\nu}-e^{-\tau(m+\kappa_p B)^2/\nu}\right].
\end{eqnarray}
By the same method for isolate poles and through a change of
variable, one arrives at
\begin{eqnarray}
\frac{\nu \beta}{8\pi^2}\lim_{\epsilon\rightarrow
0}\left\{\sum_{s,n=0}\left[\frac{(\Delta_n-s\kappa_p
B)^2}{\nu}\right]^{1-\epsilon}-\left[\frac{(m+\kappa_p
B)^2}{\nu}\right]^{1-\epsilon}\right\}\int_0^\infty dt\,
t^{\epsilon-2}\,e^{-t}.
\end{eqnarray}
The integral can be identified as $\Gamma(\epsilon-1)$. To put the
double summation in a simpler form, and bearing in mind that for
$\epsilon=0$ it reduces to
\begin{eqnarray}
\sum_{s,n=0}\frac{(\Delta_n-s \kappa_p B)^2}{\nu}=\sum_s
\frac{2\beta}{\nu}\;\zeta\left(-1,\lambda\right), \nonumber
\end{eqnarray}
where $\lambda=(m^2+\kappa_p^2B^2)/2\beta$. I make the
approximation
\begin{eqnarray}
\sum_{s,n=0}\left[\frac{(\Delta_n-s\kappa_p
B)^2}{\nu}\right]^{1-\epsilon} \approx \sum_s\;\left(\frac{2
\beta}{\nu}\right)^{1-\epsilon} \zeta(\epsilon-1,\lambda).
\end{eqnarray}
By making a Lorenz expansion in $\epsilon$ one obtains
\begin{eqnarray}
-\frac{1}{8\pi^2}\lim_{\epsilon \rightarrow
0}\Big\{\left(\frac{1}{\epsilon}+1-\gamma\right)\left[4\beta^2\zeta(-1,\lambda)-\beta\left(m+\kappa_p
B\right)^2\right]+4 \beta^2 \frac{\partial}{\partial
z}\zeta(z=-1,\lambda)\nonumber \\
 -4\beta^2 \zeta(-1,\lambda)\,\ln\left(\frac{2\beta}{\nu}\right)+ \beta\left(m+\kappa_p B\right)^2
 \ln\left(\frac{\left(m+\kappa_p B\right)^2}{\nu}\right)\Big\}.
 \end{eqnarray}
This expression has the correct limit of Eq.(\ref{EB0Lim}) for
$B\rightarrow 0$ and $\nu=m^2$. As the residue is again a
polynomial of order 4, I propose the procedure of
Eq.(\ref{ERegProc}) also for the proton case. For the properties
of the derivatives of the Hurwitz zeta function, see
\cite{ELIZALDE}.


\newpage
\begin{figure}
\includegraphics[width=0.8\textwidth]{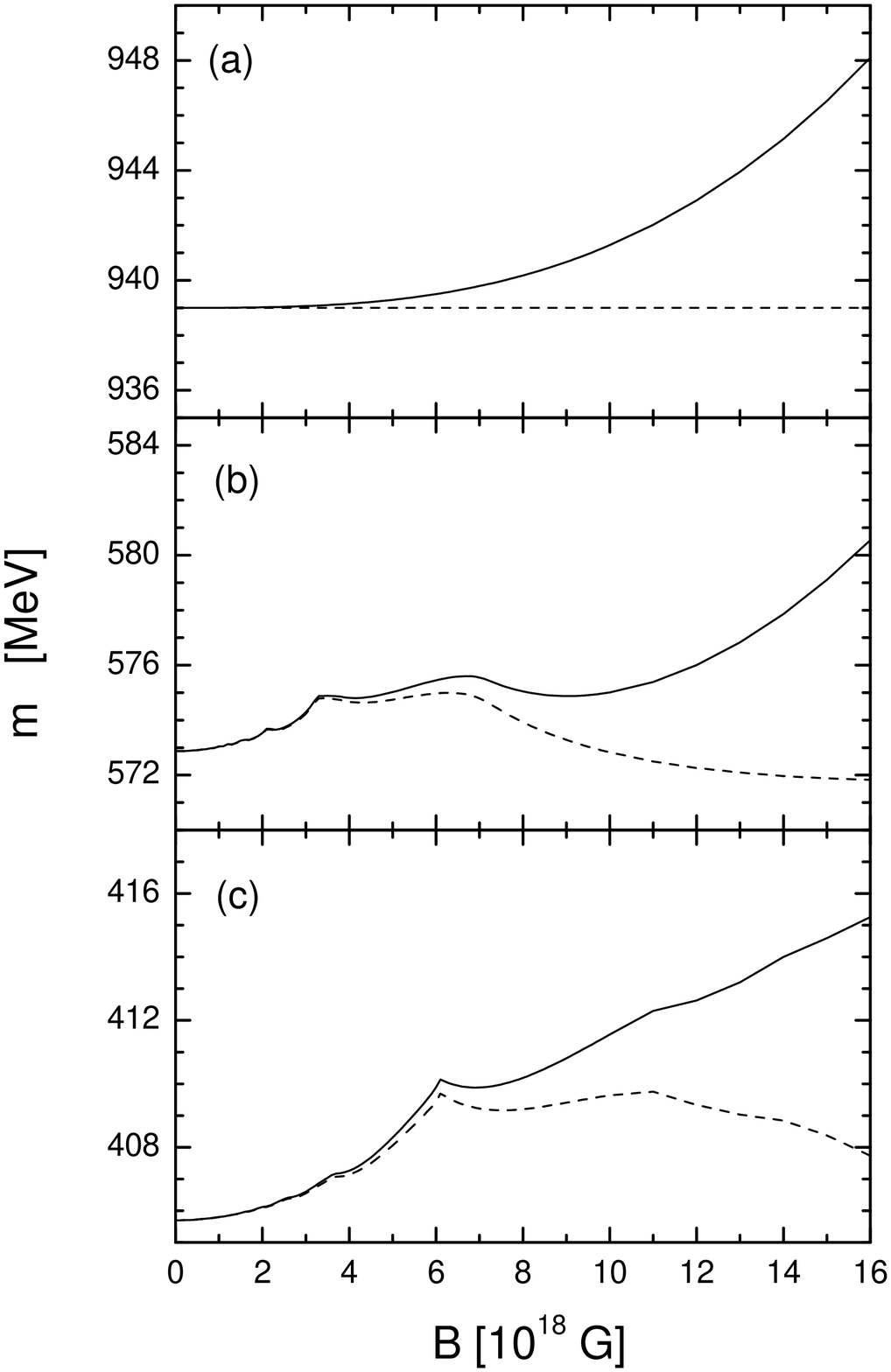}
\caption{\footnotesize The effective nucleon mass as a function of
the magnetic intensity for three different baryonic densities
$n/n_0=0$ (a), 1 (b), and 2 (c). In each case there is a
comparison of the results with (continuous line) and without
(dashed line) vacuum correction. }
\end{figure}

\newpage
\begin{figure}
\includegraphics[width=0.8\textwidth]{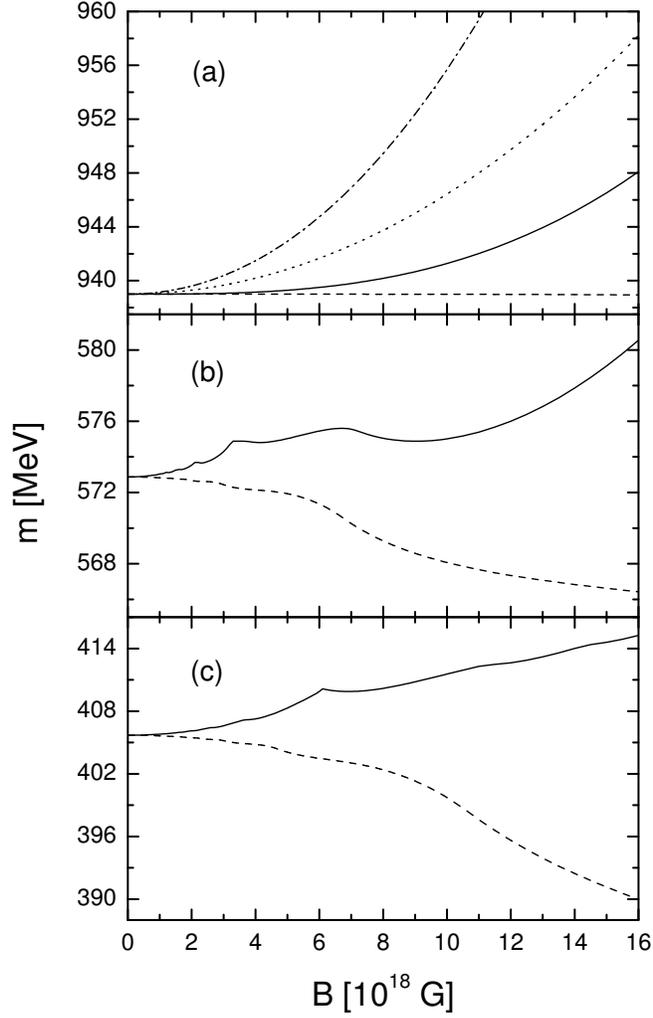}
\caption{\footnotesize The effective nucleon mass as a function of
the magnetic intensity for three different baryonic densities
$n/n_0=0$ (a), 1 (b), and 2 (c). In each case there is a
comparison of the results including (continuous line) and not
including (dashed line) the effects of the anomalous magnetic
moments. Furthermore in case (a) additional results inspired by
the calculations made in Ref.\cite{HABER} are presented, as
discussed in the main text. }
\end{figure}

\newpage
\begin{figure}
\includegraphics[width=0.8\textwidth]{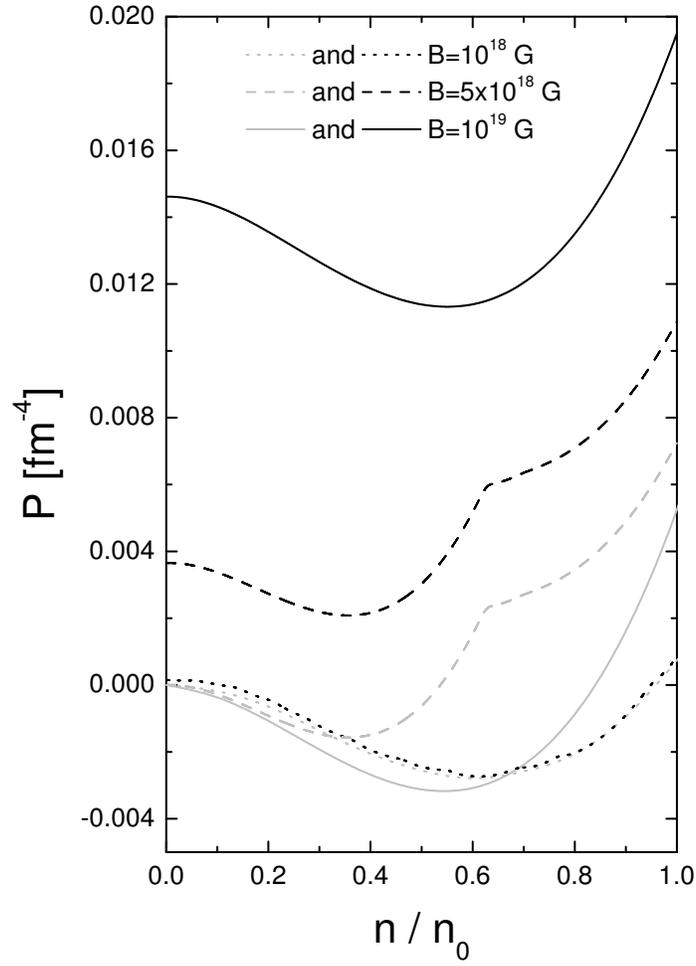}
\caption{\footnotesize The pressure as a function of the  baryonic
density for several magnetic intensities. A comparison of the
results with  and without  vacuum correction is included according
to the line convention shown in this figure. }
\end{figure}

\newpage
\begin{figure}
\includegraphics[width=0.8\textwidth]{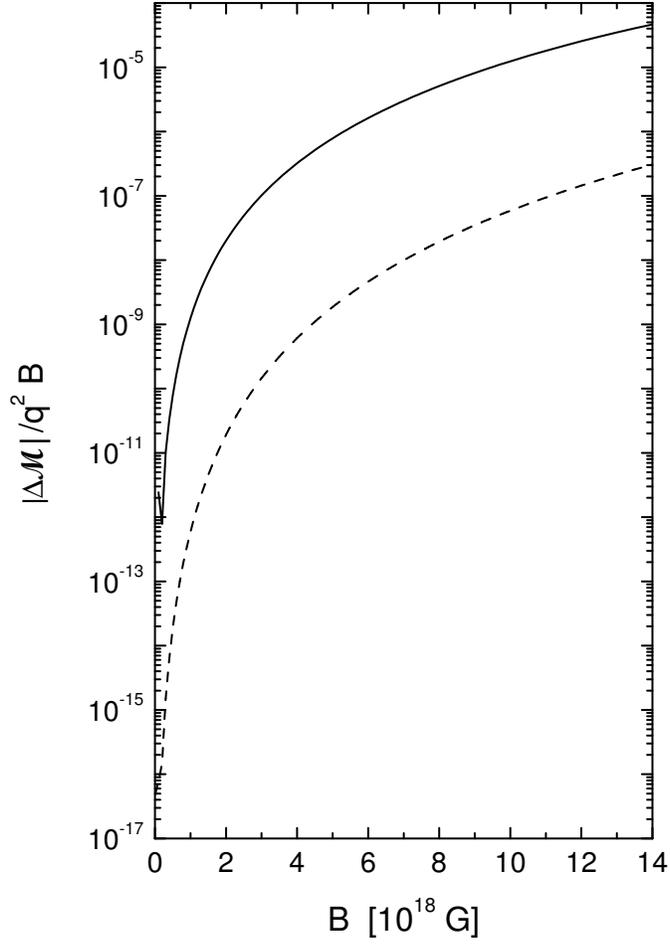}
\caption{\footnotesize The difference in the generalized
susceptibility ${\cal M}/q^2B$ between calculations with and
without vacuum corrections as a function of the magnetic
intensity. The proton (continuous line) and neutron (dashed line)
contributions have been discriminated. }
\end{figure}

\end{document}